\documentclass[%
 reprint,
 amsmath,amssymb,
 aps,
]{revtex4-2}

\usepackage{graphicx}
\usepackage{dcolumn}
\usepackage{bm}

\begin{document}

\preprint{APS/123-QED}

\title[Conductivity Freeze-Out in Isotopically Pure Si-28 at milli-Kelvin Temperatures]{Conductivity Freeze-Out in Isotopically Pure Si-28 at milli-Kelvin Temperatures}

\author{Ben T. McAllister$^{1,2,6}$}
\email{ben.mcallister@uwa.edu.au}

\author{Zijun C. Zhao$^{1,6}$}

\author{Jeremy F. Bourhill$^{1,6}$}

\author{Maxim Goryachev$^{1,6}$}

\author{Daniel Creedon$^{3,4}$}

\author{Brett C. Johnson$^5$}

\author{Michael E. Tobar$^{1,6}$}

\affiliation{$^1$Quantum Technologies and Dark Matter Laboratory, Physics Department, University of Western Australia, 35 Stirling Highway, Crawley, 6009, WA, Australia}

\affiliation{$^2$Centre for Astrophysics and Supercomputing, Swinburne University of Technology, John Street, Hawthorn, 3122, VIC, Australia}

\affiliation{$^3$School of Physics, University of Melbourne, Grattan Street, Parkville, 3010, VIC, Australia}

\affiliation{$^4$CSIRO Manufacturing, Clayton, 3168, VIC, Australia}

\affiliation{$^5$School of Engineering, RMIT University, 124 La Trobe St, Melbourne, 3000, VIC, Australia}

\affiliation{$^6$ARC Centre of Excellence for Engineered Quantum Systems, Australia, and ARC Centre of Excellence for Dark Matter Particle Physics, Australia}

\date{\today}

\begin{abstract}Silicon is a key semiconducting material for electrical devices and hybrid quantum systems where low temperatures and zero-spin isotopic purity can enhance quantum coherence. Electrical conductivity in Si is characterised by carrier freeze out at around 40 K allowing microwave transmission which is a key component for addressing spins efficiently in silicon quantum technologies. In this work, we report an additional sharp transition of the electrical conductivity in a Si-28 cylindrical cavity at around 1 Kelvin. This is observed by measuring microwave resonator Whispering Gallery Mode frequencies and Q factors with changing temperature and comparing these results with finite element models. We attribute this change to a transition from a relaxation mechanism-dominated to a resonant phonon-less absorption-dominated hopping conduction regime.  Characterising this regime change represents a deeper understanding of a physical phenomenon in a material of high interest to the quantum technology and semiconductor device community and the impact of these results is discussed.
\end{abstract}

\maketitle

\section*{Introduction}

Silicon is one of the most ubiquitous semiconducting materials in the modern world.  Its common isotopes are Si-28, Si-29, and Si-30, and it has been frequently used in the manufacture of electronic devices over the last several decades, owing to its favourable properties for such devices~\cite{Si1}. At lower temperatures, below the standard semiconductor carrier freeze-out around 40 K, the material becomes an insulator~\cite{Si2,Si3}. In more recent times, certain isotopes of silicon at low temperatures have been employed as a host for various quantum systems. 

Si-28 is commonly used for these purposes, owing to its zero nuclear spin, which leaves the medium relatively inert and decoupled from magnetic two-level systems, greatly reducing the available loss mechanisms in the material. Consequently, Si-28 can be an excellent host for well-isolated spin-based quantum systems, such as qubits~\cite{HQS1,HQS2}. Furthermore, there are various proposals to employ Si-28 in next generation solid state clocks~\cite{Clocks1,Clocks2}. Many of these applications rely on interfacing Si-28 with microwave radiation, and coupling samples of Si-28 to microwave resonances.  It is important to understand the various loss mechanisms in materials used in quantum systems, and for losses to be as low as possible in order to maximise coherence times~\cite{LossTangent1, LossTangent2, LossTangent3}. Therefore there is a high premium placed on understanding the electrical properties of silicon, particularly isotopically pure Si-28 at low temperatures, and at microwave frequencies for applications in quantum technology, where material losses in substrates are currently one of the key limiting factors~\cite{SiLosses1,SiLosses2,SiLosses3,SiLosses4,SiLosses5,SiLosses6}. 

Previous work has investigated silicon under these conditions. A common method for conducting such investigations is the established technique of exciting electromagnetic Whispering Gallery Modes (WGM) inside samples of material~\cite{WGM1,WGM2,WGM3,WGM4}, and using the resonant mode properties such as frequency and Q factor to derive the host material properties such as permittivity, and loss. This has been performed previously for Si-28 at 4 K and mK temperatures~\cite{Krupka,Bourhill,Kostylev}. These studies reported the highest Q factors of such modes at milli-Kelvin temperatures, on the order of $10^6$ at microwave frequencies. The authors proposed that the losses in the $\sim10$ GHz range are limited by hopping conduction~\cite{Hopping1,Hopping2}, and dielectric losses, but were not able to disentangle these factors. 

In silicon, there exists a standard semiconductor freeze-out below 40 K due to dopant electrons being unable to escape from the valence band to the conduction band. However, some residual conductivity remains due to a range of mechanisms that are referred to as hopping conductivity. Various hopping and other conduction mechanisms at low temperatures in semiconductors have been proposed and observed~\cite{Si4New}. Generally, hopping conduction refers to conduction in a semiconductor enabled by the `hopping' of electrons between individual host sites~\cite{Si3}.

\begin{figure}
    \centering
    	\includegraphics[width=0.95\columnwidth]{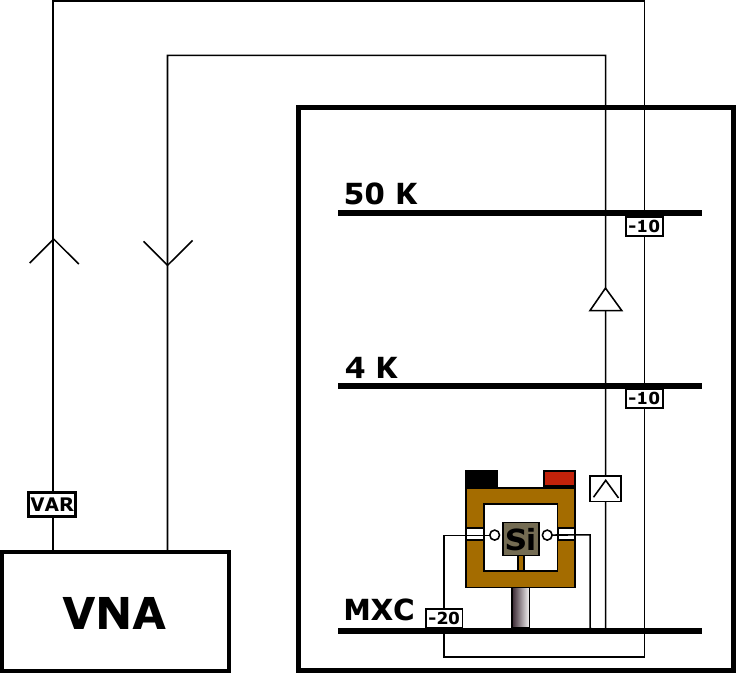}
    \caption{A schematic of the measurement setup. The cavity is mounted to the mixing chamber (MXC) plate of a dilution refrigerator, coupled to via loop antennae, and read out with cryogenic and room temperature amplification. Various attenuation is distributed along the input line, including a variable attenuator at room temperature (VAR) and 10 and 20 dB attenuators at various other stages (-10, -20). The location of the resistive heater is represented by the black box, whilst the location of the temperature sensor is represented by the red box. The cylinder coloured with the grey gradient represents the niobium superconducting link to the MXC plate. Isolation is represented by the box containing an arrow, and amplification is represented by a white triangle.}
    \label{fig:schem}
\end{figure}

More recent work has made detailed measurements of the low temperature dependence of the conductivity in high resistivity (but not isotopically pure) silicon~\cite{Checchin}. This study observed conductivity reducing with temperature, and observed the presence of additional residual conductivity above the hopping conduction predicted by their model, with the additional residual conductivity dominating at low temperatures. The sample in question appeared to exhibit overall slightly higher losses than the previous studies in isotopically pure silicon, which can perhaps be accounted for by this residual conductivity. As such, further detailed study of the hopping conduction regime is well motivated in isotopically pure silicon, for the sake of further understanding an important material in quantum technologies.

In this work, we study the temperature dependence of material losses in isotopically pure Si-28, with a particular focus in the region around $\sim 1$ K and at frequencies in the 10s of GHz. We use a similar technique as Bourhill et al.~\cite{Bourhill}, but with a variable temperature for the Si-28 resonator. We find a sharp dependence of the Q factor of the resonances on the temperature of the sample around $\sim 1$ K, followed by a plateau.  We discuss candidate mechanisms which could be responsible for this behaviour, and show that it is consistent with a freeze-out - specifically a transition from the hopping conduction regime dominated by the phonon-assisted relaxation mechanism, to the regime dominated by resonant phononless absorption of individual photons. The transition temperature between these regimes is expected at temperatures $k_BT<hf$, which corresponds to temperatures on the order of 1 K for conduction frequencies on the order of 10 GHz. This result shows a limit to the decrease in conductivity that can be achieved by simply cooling the material further, and is qualitatively consistent with other recent results~\cite{Checchin}.

In the region where $k_BT>hf$ and where the hopping conduction is dominated by phonon-assisted electron tunnelling (the relaxation mechanism), theory predicts that the dependence of the conductivity, $\sigma$ is given by~\cite{Si4New}

\begin{equation}
\sigma(T,f)\propto\sigma_0\times T\times f(ln(\nu_p/f))^4.
\label{eq:relax}
\end{equation}

Here $\sigma_0$ is some constant, $T$ is the temperature, $f$ is the frequency of conduction, and $\nu_p$ is the phonon frequency, typically of order $10^{12}$ Hz~\cite{Si4New}. In the region where $k_BT<hf$, where theory predicts conduction is dominated by resonant phononless absorption of photons enabling electron hops, the conductivity is given by

\begin{equation}
\sigma(T,f)\propto\sigma_0\times f.
\label{eq:phononless}
\end{equation}

So, as the sample changes temperature and the conduction transitions from one regime to the other, we expect to see the disappearnce of temperature dependence. This transition has not been experimentally characterised before, to our knowledge.

\section*{Experiment}

\begin{figure*}
    \centering
    \includegraphics[width=0.95\textwidth]{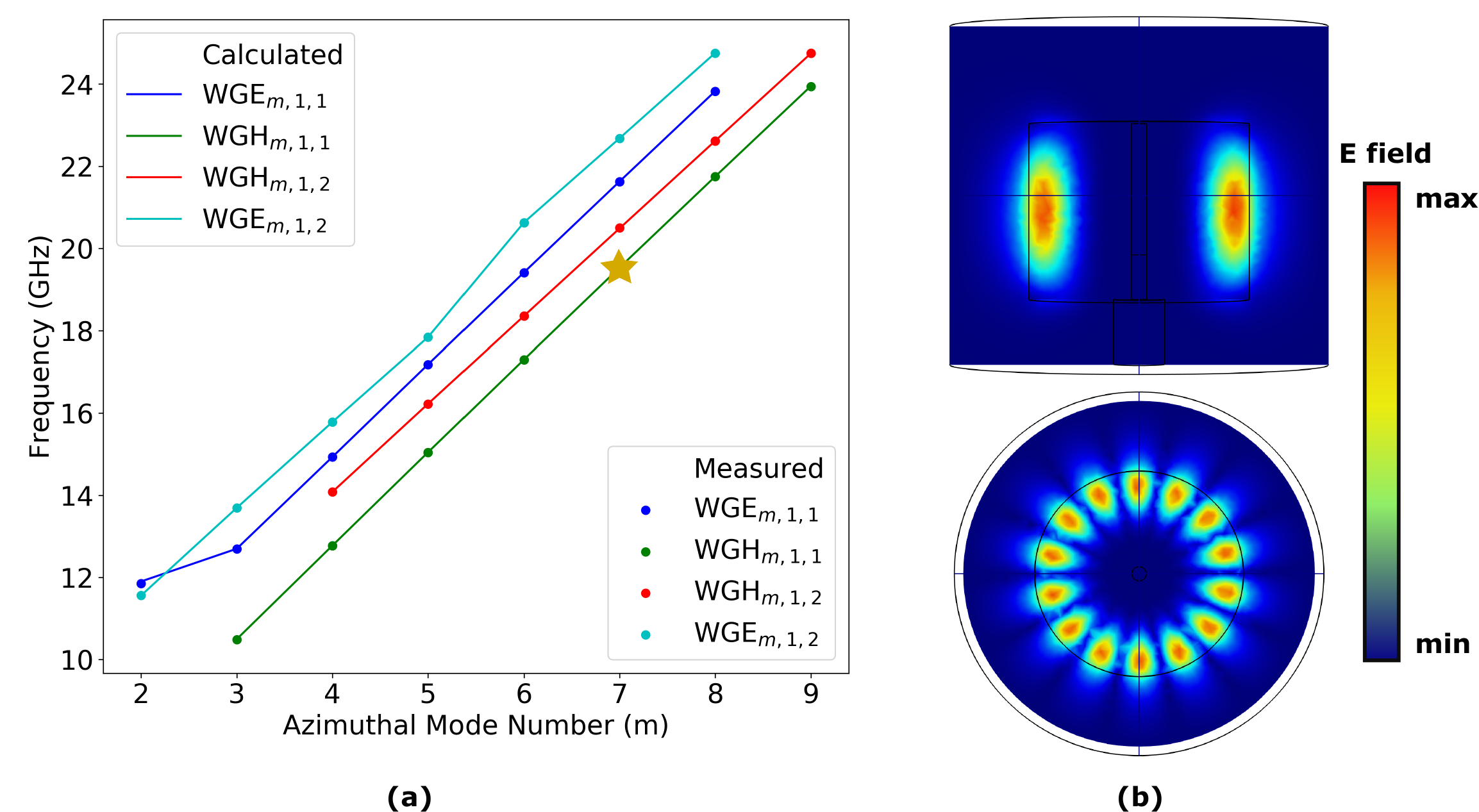}
    \caption{Results of the finite element modelling: (a) calculated vs measured frequencies for WGH and WGE mode families (b) two views of the normalized electric field magnitude for the WGH$_{711}$ mode, on top is a side profile, on the bottom is a top-down view. This mode is indicated by the gold star in (a).}
    \label{fig:modes}
\end{figure*}

Our experiment consists of an isotopically pure Si-28 resonator precisely machined into a uniform cylinder from a larger boule. The machined cylinder was mechanically polished and underwent an annealing, oxidation, and passivation protocol~\cite{Bourhill} to remove surface impurities and strain, and minimise losses from fixed oxide charge and surface dangling bonds. In particular the cylinder has been mechanically polished and annealed, to remove surface impurities and associated losses.  In the experiment, the sample is placed inside a copper microwave resonant cavity, and coupled to with loop antennae. This system is cooled inside a refrigerator below the silicon semiconductor freeze-out at 40 K. In this way, we are able to excite hybrid silicon WGM - copper cavity modes. Detailed information on such systems and their experimental design can be found in~\cite{WGM1,WGM2,WGM3,WGM4}. The model for the total Q factor, $Q_T$ of the hybrid electromagnetic resonances which exist in this system is given by~\cite{Bourhill, Kostylev, NbPost}
\begin{equation}
\frac{1}{Q_T} = \frac{R_s}{G} + p_e\tan{\delta} + p_e\frac{\sigma}{\omega\epsilon_0\epsilon_r} + \frac{1}{Q_{\text{external}}}.
\label{eq:QT}
\end{equation}
Here, $R_s$ is the surface resistance of the copper walls of the outer cavity, $G$ is a geometry factor, relating to the field profile of the mode in question, $p_e$ is the electrical filling factor, a measure of the amount of the electric field of the resonance which is localised inside the silicon sample,  $\tan\delta$ is the dielectric loss tangent, $\sigma$ is the conductivity of the silicon sample, $\omega$ is the mode angular frequency, $\epsilon_0$ and $\epsilon_r$ are the vacuum and relative permittivity of the silicon sample, and $Q_\text{external}$ account for other losses, such as losses due to the antenna coupling, or radiation leakage. $G$ and $p_e$ are given by
\begin{equation}
    G = \frac{\omega\mu_0\int_{V_t} dV \vec{H}^2}{\int_{S_c} dS \vec{H}^2},
\end{equation}
and
\begin{equation}
    p_e = \frac{\int_{V_s} dV \vec{E}^2}{\int_{V_t} dV \vec{E}^2},
\end{equation}
where $\mu_0$ is the vacuum permeability, $\vec{H}$ is the mode magnetic field, $\vec{E}$ is the mode electric field, $V_s$ is the volume of the silicon sample, $V_t$ is the total volume, and $S_c$ is the conducting surfaces of the copper cavity resonator.

We model the system via the finite element method using COMSOL Multiphysics, computing $G$, $p_e$, and compare with experimental Q factor data to extract material properties.

Figure~\ref{fig:schem} is a schematic of the experiment. A cylindrical copper resonant cavity with a 12.33 mm radius and 22.95 mm height was employed. The cavity consisted of a hollow cylindrical shell with two cylindrical endcaps. Losses from the seams were minimised by use of knife-edge contacts and, as we will see the modes of interest are highly localised inside the silicon sample. The Si-28 WGM resonator had a 7.23 mm radius and 11.9 mm height, with a small hole in the base, so that it could be supported on a 1.68 mm radius sapphire post in the base of the cavity. Loop antennae were inserted into the sides of the copper cavity, oriented at 45 degrees to the horizontal so as to excite both transverse electric (WGE) and transverse magnetic (WGH) modes. The antennae were positioned so that they were only slightly penetrating the cavity volume, to minimize coupling and thus loading of the mode Q factors by the external circuit. The cavity containing the Si-28 resonator was bolted to the mixing chamber plate of a Bluefors LD250 dilution refrigerator, and a resistive heater and temperature sensor were installed on the cavity, as per the schematic.

The loop antennae were connected to cryogenic microwave cables that were installed in the dilution refrigerator, and various other microwave components were employed. A low noise HEMT amplifier at the 4 K stage was used to improve signal to noise, with an isolator between the cavity and the amplifier to reduce the propagation of amplifier noise back into the resonator. Attenuation was employed at various stages to reduce the propagation of room temperature thermal noise into the experiment, and to reduce the input power to the experiment and avoid heating the sample with excess microwave power. The refrigerator was connected to a Vector Network Analyzer (VNA) by room temperature microwave cables, and additional room temperature HEMT amplification was used as needed to boost signal to noise. Injected powers from the VNA were on the order of 0 dBm, but heavily attenuated. A variable amount of room temperature attenuation on the order of 30-40 dB was employed, followed by 40 dB of cold attenuation distributed along the input line. Additionally, the microwave cables and connectors at room temperature and inside the cryogenic system have been shown to exhibit 10-20 dB of loss in previous measurements. With the low coupling to the resonator, it is estimated that less than -100 dBm of microwave power is delivered to the sample.

A novel aspect of this experimental setup is the presence of a niobium linking piece between the cavity assembly, and the MXC plate. The purpose of this was to allow for enhanced isolated temperature control of the cavity assembly, without heating the dilution refrigerator directly. Previous iterations found that when heating the cavity to temperatures near 1 K, the refrigeration system would begin to struggle with the heat load, greatly limiting the range of temperatures that could be studied to be between the base temperature of the refrigerator ($\sim 20$ mK) and about 700 mK. When the experiment is cooled below the critical temperature of niobium the linking piece becomes superconducting, and thus serves as a very poor thermal link, creating isolation. In this mode of operation, the cavity did not reach the same ultimate base temperature, instead thermalizing to about 100 mK, but it was possible to heat the cavity to nearly 8 K before the heat load became too great for the refrigerator, allowing for a much wider range of temperature data to be taken within the same experimental run. The piece used was a simple rod of niobium which was bolted to the mixing chamber plate, with a cylindrical platform at the upper end, which the cavity was mounted to with screws. This technique may be readily applied to other similar measurements, and can be thought of as a `natural' cryogenic heat switch mechanism which kicks in below the critical temperature of the chosen material. Of course, choosing materials other than niobium with different critical temperatures could allow for different ranges of operation. The technique may be particularly useful in characterising superconducting materials with critical temperatures in the range of a few Kelvin, where it is often difficult to heat dilution refrigerator base plates without overloading the system.

\begin{figure*}
    \centering
    \includegraphics[width=0.95\textwidth]{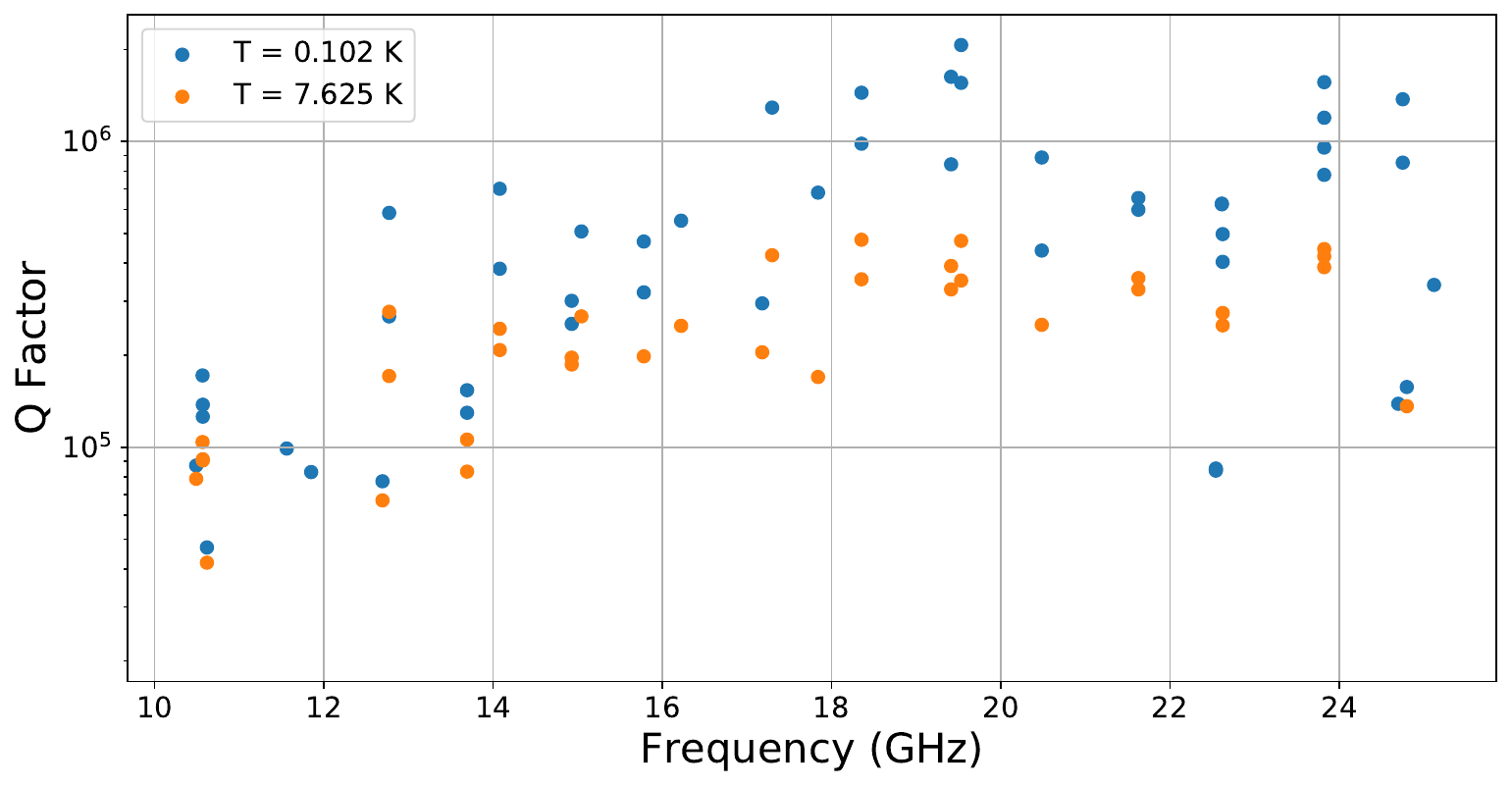}
    \caption{Q factor vs frequency for the modes studied. Blue dots represent the base temperature of 102 mK, and orange dots represent the highest temperature of 7625 mK. As the temperature fluctuated somewhat during the frequency sweeps, the temperatures reported for each series of measurements correspond to the mean of the temperatures recorded on the sensor during the period of measurement. On this scale it may appear that there are multiple data points at the same frequency, but there is only a single yellow and a single blue dot for a given mode. The mode spacings are sometimes on the order of a few MHz. This plot serves to show the general trend in increased Q factor between the extremes of the temperature range, and the frequency dependence of the effect. As discussed below, uncertainty in measured and fitted quality factors should be taken to be $\sim10\%$, with error bars omitted here for readability.}
    \label{fig:Qvsf}
\end{figure*}

Once the cavity reached its base temperature, spectroscopy was performed with the VNA to observe microwave modes in the 10 - 30 GHz range. These modes were matched in frequency against modes from a COMSOL finite element model, as shown in Fig.~\ref{fig:modes}.

The COMSOL model employed the electromagnetic waves physics package, and performed eigenfrequency analysis on a model constructed with the geometries outlined above. The permittivity value for the silicon was 11.488~\citep{Bourhill}, and the tetrahedral mesh consisted of roughly 350,000 mesh elements.

Once high-Q modes were identified, a power dependence sweep was performed to determine the optimal input microwave power level in order to maximise the Q factor of the resonances at the base temperature. Ultimately, it was found that injecting the optimal amounts of power provided too high a heat load, and caused the sample to heat in an unstable fashion. As such, a compromise power was chosen at each frequency, which achieved a Q factor within $\sim$5\% of the maximum Q factor at base temperature, but which minimized temperature instability of the sample.

After an initial run at mK temperatures, the cavity was warmed, and the antennae were slightly removed from the cavity to reduce the loading of the resonances by the external circuit to ensure that we are in the vastly under-coupled regime ($\beta<<10^{-2}$), and verify that the material properties were the dominant loss mechanism~\cite{NbPost}. After re-cooling the setup, the Q factor did not change substantially with this lower coupling, and so we deem Q factor loading by the external circuit to be negligible. Radiation losses were deemed to be insignificant, with finite element modelling of the system indicating near-perfect confinement of the modes within the silicon resonator, with filling factors near unity ($p_e>0.94$ for all modes studied).

Spectroscopy data was taken with the VNA for 47 high-Q microwave modes at the base temperature. Various other modes with Q factors below a few tens of thousands were excluded from the study, as with radiation, surface impurities, or copper wall losses as the dominant loss factor, it would be more difficult to resolve changes in the hopping conduction. As such, only the highest-Q modes were selected for study. The cavity was then successively heated by applying current to the resistive heater mounted to the cavity, and a period of approximately 30 minutes was allowed to pass after each successive heating, to allow the temperature as measured on the temperature sensor to stabilize, and to ensure good thermalization across the entire resonator/silicon assembly. Spectroscopy data was taken with the VNA at 9 temperatures, up to a maximum of approximately 7.6 K, following the procedure outlined above. As the temperature was increased, all other experimental parameters were left constant, to ensure that any changes were a result of temperature induced effects.

Similar data were taken in earlier runs at lower temperatures, without the niobium linking piece. Whilst the temperature behavioural trend in Q factors in the 20 mK base runs and the 100 mK base runs were consistent, there were discontinuities in Q vs temperature when attempting to `stitch' the two data sets together, and so we decided to use only data from a single experimental run in a given analysis. The niobium isolated run was chosen, since it afforded a much wider range of temperatures within the same cool and increased confidence in the thermalisation of the temperature sensor with the cavity and sample, despite a slightly higher base temperature.

\section*{Analysis}

Once the data were acquired, a model was fit to the S21 transmission spectra of the resonances, to determine the Q factor of each mode at each temperature. Due to the asymmetry in the background, we employed a Fano resonance~\cite{Fano1,Fano2} fit, with the profile
\begin{equation}
    S_{21} = k\times(1-\frac{(q\Gamma_{r}/2+f-f_0)^2}{(\Gamma_{r}/2)^2+(f-f_0)^2}).
\end{equation}
Here, $k$ is an overall loss factor to account for the background losses and gain in attenuation, amplification, cables etc, $q$ is a Fano fitting paramter, related to the degree of asymmetry in the fit, $\Gamma_r$ parameterizes the line-width of the resonance, $f$ is frequency, and $f_0$ is the central frequency of the resonance. The Q factor is extracted from this fit, given by $Q=f_0/\Gamma_r$.

\begin{figure*}
    \centering
    \includegraphics[width=0.95\textwidth]{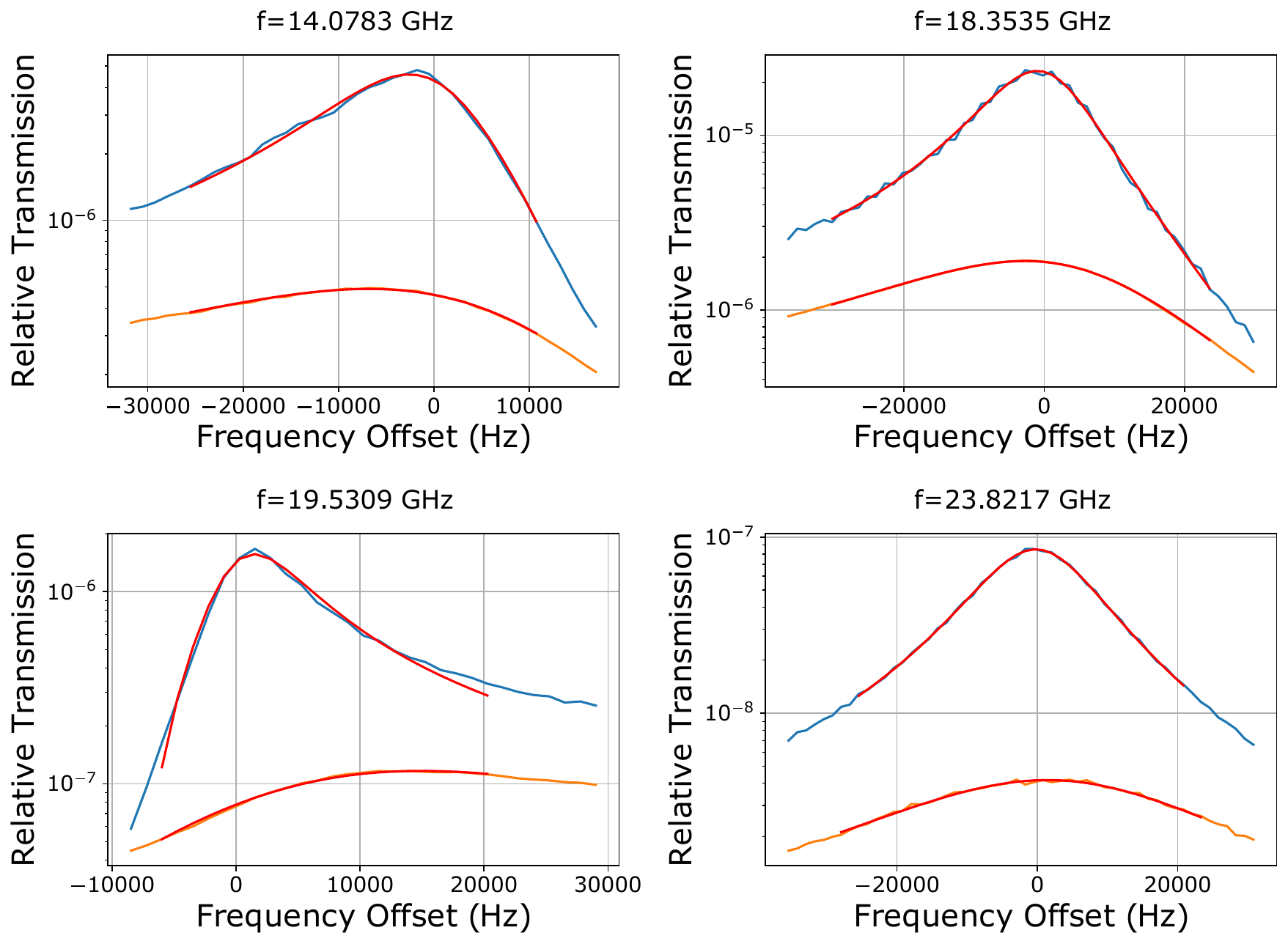}
    \caption{Relative transmission data against frequency in linear units for 4 selected modes of the 47 studied. Blue lines represent the magnitude of the experimental transmission data at the base temperature of 102 mK. Orange lines represent the magnitude of the experimental transmission data at the highest temperature of 7625 mK. Red lines are truncated Fano fits applied to extract Q factors. The modes were identified from finite element modelling as follows. Top left: WGH412, top right: WGH612, bottom left: WGH711, bottom right: WGE811. }
    \label{fig:Qfit}
\end{figure*}
\begin{figure*}
    \centering
    \includegraphics[width=0.95\textwidth]{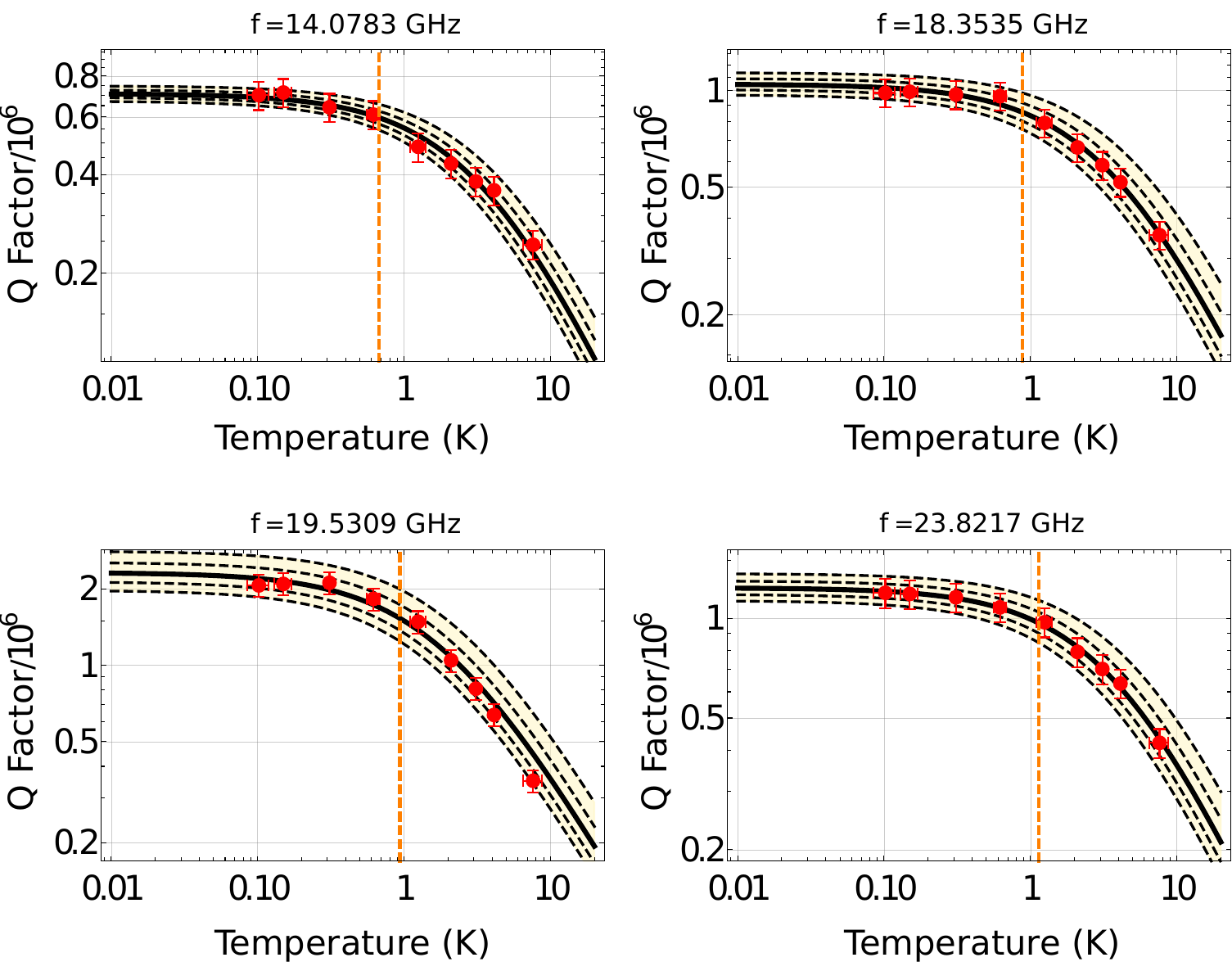}
    \caption{Q factor versus temperature data for the 4 selected modes shown in Fig.~\ref{fig:Qfit}. Q factor values derived from Fano fits are shown with the red dots. The uncertainty in Q factor values is taken to be $\sim$10\% in line with typical previous experiments~\cite{QVNA}.  The solid black lines show the central value of Q factor as a function of temperature which was fit to the experimental data (red dots) with the model presented in eq.~\eqref{eq:QvsT}. It is from these fits that conductivity was derived. The dashed black lines represent the one and two sigma error bars on these fits, whilst the dashed orange line represents the temperature at which the transition is expected to occur according to the model. The plots for the remaining modes used in fitting are displayed in the Appendix.}
    \label{fig:QvsT}
\end{figure*}

This fitting procedure was applied to each of the 47 high-Q resonances at each of the 9 studied temperatures, to extract Q factors. In some lower-Q modes at the higher temperatures, the signal to noise ratio became too low to achieve an adequate fit, and these data were removed from the study. Fig.~\ref{fig:Qvsf} shows the frequency dependence of the Q factors measured, at the base temperature, and the highest temperature studied. The best Qs found are on the order of $10^6$, in-line with but slightly higher than previous work~\cite{Bourhill}. It has been found in other studies that the uncertainty in Q factors from uncalibrated VNA traces fit in transmission, such as these, can be approximated at 7\%~\cite{QVNA}, in the interest of being conservative, we estimate the fitted Q factor uncertainties here at 10\%.

In general, it was found that over the frequency range 10 - 30 GHz, the high-Q mode Q factors increased as the temperature lowered, until a plateau was observed below 1 K. Transmission spectra for 4 selected modes at different positions across the frequency range which are exemplary of this behaviour are shown, along with the Fano resonance fits in Fig.~\ref{fig:Qfit}. The temperature dependence of the Q factors for these same 4 modes are shown in Fig.~\ref{fig:QvsT}. In general, the observations support the idea of a transition at $k_BT < hf$.

With Q factor values in hand, the next step was to extract conductivity values using the model presented in \eqref{eq:QT}, in conjunction with assumed values for some parameters, and inputs from the finite element modelling for others. 

The process of extracting conductivity values was as follows. First, we selected 10 modes, a mixture of WGH and WGE modes with high Q factor values at base temperature, on the order of $10^6$. High Q factor values are required to resolve small changes in conduction. A mode which has dominant loss factors in the form of radiation, or resistive wall losses will be difficult to extract useful information from. These modes ranged in frequency from 14 - 24 GHz, and include the 4 modes presented in figs.~\ref{fig:Qfit} and~\ref{fig:QvsT}. For each of these modes, we constructed a list of Q factors as a function of temperature, similar to the data presented in Fig.~\ref{fig:QvsT}. We used the central Q values derived from the Fano model fitting for each, understanding as above that there is an uncertainty of $\sim 10\%$ on these values. We then made a model of the Q factor as a function of temperature, derived from the model presented in~\eqref{eq:QT}. This model supposes that
\begin{equation}
Q(T,f) = \left(\frac{\sigma(T,f)}{2\pi f\epsilon} +  \frac{1}{Q_{\text{other}}}\right)^{-1}.
\label{eq:QvsT}
\end{equation}
Here $Q_{\text{other}}$ absorbs all non-silicon conductivity-related loss terms (radiation, dielectric, resistive walls) which are assumed to be constant over the temperature range of study (100 mK to 8 K). We consider this a safe assumption, as it is known that the resistivity of copper does not change substantially below a few K, nor do the dielectric properties of silicon~\cite{Bourhill}, and there is no known reason that radiation losses should change over this range, as the geometry of the resonator is not changing. Furthermore, the modes are highly confined inside the silicon sample, and wall losses and radiation should not be the dominant effects. As such, the only quantity expected to change is the silicon sample conductivity. Given we expect the conductivity's temperature and frequency dependence to change from the form given in~\eqref{eq:relax} to that given in~\eqref{eq:phononless}, for the purpose of fitting we model conductivity with a power law addition of the two forms, where the temperature and frequency dependence change at $k_BT=hf$. The frequency dependence of~\eqref{eq:relax} is very shallow in the frequency range of interest, amounting to a change of less than 1\% in conductivity as frequency changes from 14 to 24 GHz, at a fixed temperature. So, we model the conductivity with a function which changes from having a temperature dependence of $T^1$ to $T^0$, and frequency dependence $f^0$ to $f^1$, with the change occurring at $k_BT<hf$. We model the conductivity with a fitting function given by
\begin{equation}
\sigma(T,f)=\sigma_0\frac{hf}{k_b}+\sigma_0 T.
\label{eq:sigmaT}
\end{equation}
Here $\sigma_0$ is some constant which sets the conductivity. This is the only free parameter in the conductivity model, and the only other free parameter in the Q vs temperature model is $Q_\text{other}$. We use a linear addition power law to govern the transition between regimes - higher and lower power laws were considered, which alter the sharpness of the transition, but were ultimately found to have negligible impact on the fitted conductivity values (within uncertainty). Thus for simplicity, a linear addition was used. We have assumed the electrical filling factor $p_e\approx 1$, which was the result of the COMSOL modelling for all of the high-Q modes. We note that this model has the transition temperature of $k_BT<hf$ `built-in', along with the correct dependence on $T$ and $f$ in each temperature region. 

We then fit~\eqref{eq:QvsT} to each of the 10 data sets, to extract $\sigma_0$ and $Q_\text{other}$. $Q_\text{other}$ should be mode dependent, $\sigma_0$ should not be. For an initial guess of $Q_\text{other}$, we used the highest Q value from each data set, but left it as a free parameter, and recorded the fitted value.

\section*{Results and Discussion}
\begin{figure*}
    \centering
    \includegraphics[width=0.95\textwidth]{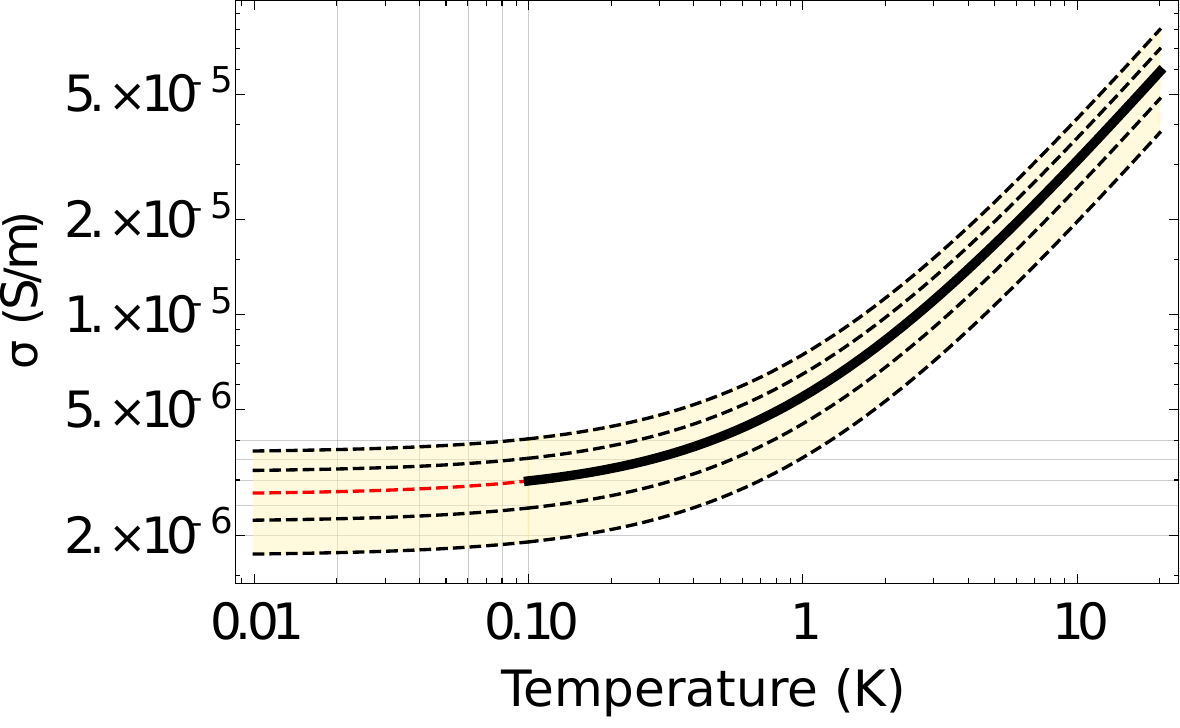}
    \caption{Fitted value of conductivity against temperature at 20 GHz, using the parameters averaged across the 10 high Q factor modes chosen. The solid black line represents the central value, and the dashed black lines represent the one and two sigma error bars on the fit. The region where the solid black line transitions to a dashed red line represents the region where experimental data were not taken, and is therefore an extrapolation.}
    \label{fig:sigmavsT}
\end{figure*}

After fitting we arrived at a mean value for $\sigma_0=2.81\times10^{-6}$ S/m, with a standard deviation of $0.50\times10^{-6}$ S/m, and a mean value of $Q_\text{other}=2.16\times10^{6}$, with a standard deviation of $1.40\times10^6$. We re-iterate that $Q_\text{other}$ should not be considered a global value, but a mode dependent one - the mean value is useful only as an estimate of the order of magnitude of radiation and other losses. We have included all 10 Q factor against temperature fits in the Appendix. A subsequent plot of $\sigma(T)$ at $f=20$ GHz, using the averaged fitted values is shown in Fig.~\ref{fig:sigmavsT}. The uncertainties in measured Q factors and temperatures were factored into the fitting routine via a multiple least squares method, and the resulting standard errors from each of the 10 individual fits were computed. However, when considering the distribution of the central $\sigma_0$ values from these 10 fits (from which the mean value was computed) the standard deviation of the fitted value of $\sigma_0$ is greater than standard error from the fitting algorithm (close to 18\% of the central value). In the interest of being conservative, we deem the fitting errors to be too optimistic, and we take the uncertainty in the $\sigma_0$ to be on the order of the standard deviation of $\sigma_0$ values extracted from fitting. Considering~\eqref{eq:sigmaT}, we can see that $\delta\sigma/\sigma=\delta\sigma_0/\sigma_0$, and thus the uncertainty in the extracted values of $\sigma$ is also taken to be approximately 18\%, as reflected in the plot.

Due to the agreement of the model with the data as shown in Fig.~\ref{fig:QvsT} we conclude that we do indeed observe a transition from the regime where conductivity is dominated by the relaxation mechanism, to a regime where the temperature dependence freezes out, and electron hopping conduction is dominated by absorption of photons. Importantly, this transition has not been characterised in detail before, to our knowledge.

Critically, these results indicate that there are diminishing returns to be achieved in terms of microwave losses in isotopically pure Si-28 below 1 K, meaning that it may not be necessary to cool quantum devices hosted in Si-28 to the lowest temperature to achieve the lowest microwave losses. For instance, if quantum devices hosted in silicon were to be developed at higher frequencies (around 20 GHz, as opposed to the 4-8 GHz band which is common) one could potentially avoid the need to cool these devices to deep milliKelvin temperatures to ensure that the thermal occupation of photons was minimal, without incurring additional substrate losses owing to higher temperatures. This could apply to various systems, including spin-based and superconducting qubits hosted in silicon, where microwave radiation is used to address the system and material losses may play a role. This observation could simplify experimental complexities for such experiments, and is in qualitative agreement with other recent work~\cite{Checchin}.
 
It is difficult to explain these results through standard dielectric losses, copper surface resistances losses, or coupling losses, and given the modelled conductivity obeys the expected temperature dependence for the hopping conduction regimes discussed, we regard this explanation as highly likely.  However, it is also possible some additional exotic loss mechanism, such as charge carriers on the surface of the sample due to impurities undergoing a change of regime could be responsible for this behaviour.  We consider this unlikely given the high purity of the isotopically pure Si-28 sample in question, as determined previously~\cite{Kostylev}, and the fact that the effect persists in both WGE and WGH mode families, with different degrees of coupling between the electromagnetic radiation of the mode and the surface of the sample.

It must also be noted that previous measurements have shown a similar transition in conductivity at much higher temperatures (10s of K)~\cite{Si4New}. This effect is attributed to electron-electron interactions, rather than the freeze-out of the relaxation mechanism. If we are witnessing the same effect here, it is remarkable that it happens at such low temperatures, and temperatures consistent with the $k_BT < hf$ transition temperature as predicted by the resonant phononless absorption regime. For this reason we consider it more likely that we are observing the transition to the phononless absorption regime - however, in either case, the final result on the conductivity is qualitatively the same. There are diminishing returns in cooling below a certain low temperature, which may simplify experimental configurations for quantum technology.

\section*{Conclusion}

Microwave losses in isotopically pure Si-28 have been studied at low temperatures, and as a function of frequency and temperature. A clear transition in the dominant loss mechanism takes place at a temperature near 1 K, which we show can be explained by a transition in the regime of hopping conduction in the sample. Whilst hopping conduction has been observed and characterised in silicon before, this transition has not been experimentally reported in the literature to our knowledge. This is of broad interest to the field of quantum systems engineering, as Si-28 is increasingly attractive as a host for qubits, and for potential applications in quantum clocks. We employed a WGM resonator at low temperatures combined with data analysis techniques, and finite element modelling to extract the material parameters.

\begin{acknowledgments}

This work was supported by The Forrest Research Foundation, and Australian Research Council Centre of Excellence for Engineered Quantum Systems (CE170100009) and Centre of Excellence for Dark Matter Particle Physics (CE200100008). The authors would like to acknowledge useful discussions with Boris Shklovskii.

\end{acknowledgments}

\newpage~\newpage
\section*{Appendix}\label{secA1}
As discussed in the text, we will now present the Q factor versus temperature plots and fits for all 10 utilized modes.  All plots follow the same conventions as the corresponding plots in the text.

\centering
\includegraphics[width=0.95\columnwidth]{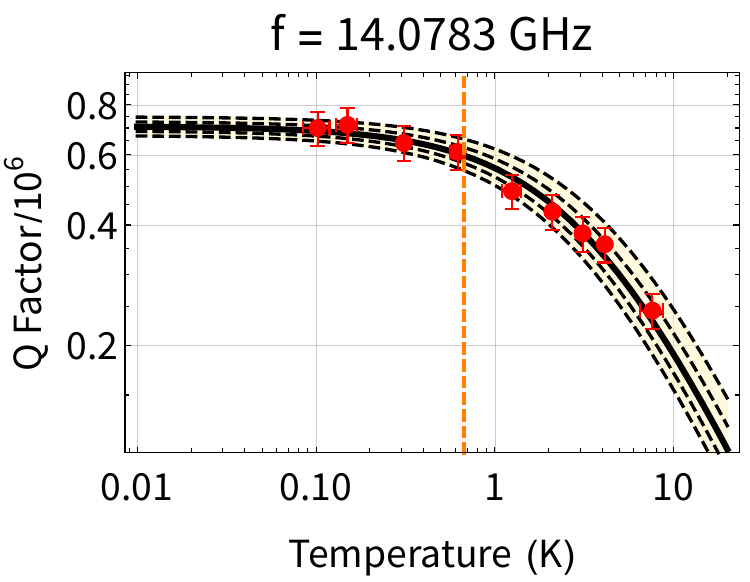}

\centering
\includegraphics[width=0.95\columnwidth]{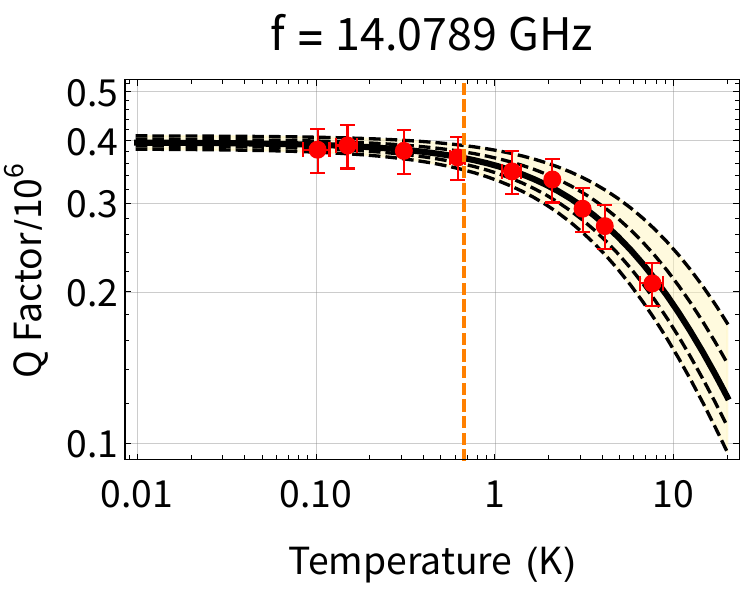}

\centering
\includegraphics[width=0.95\columnwidth]{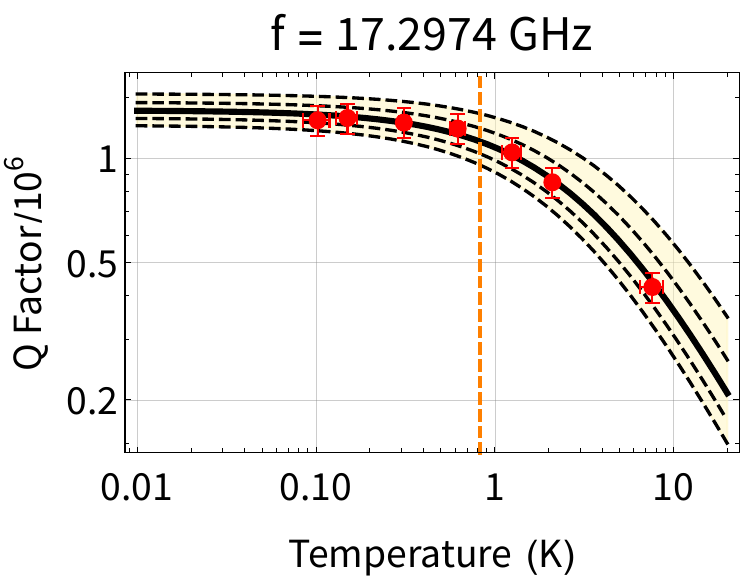}

\centering
\includegraphics[width=0.95\columnwidth]{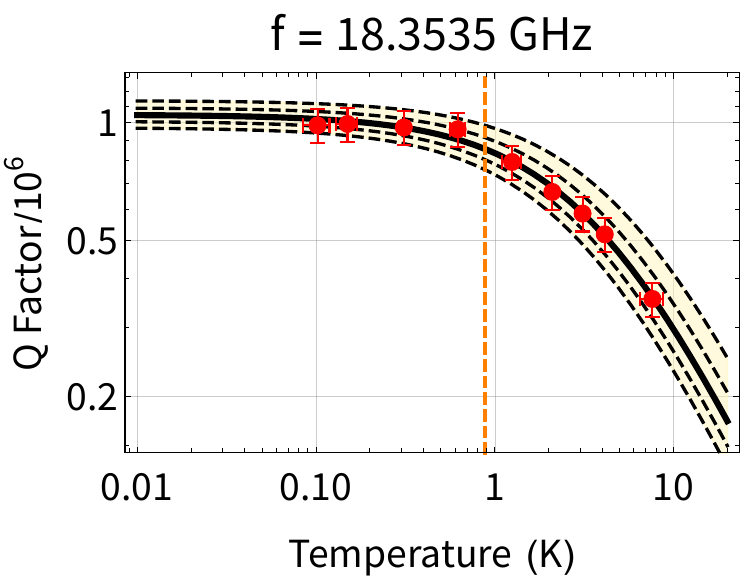}

\centering
\includegraphics[width=0.95\columnwidth]{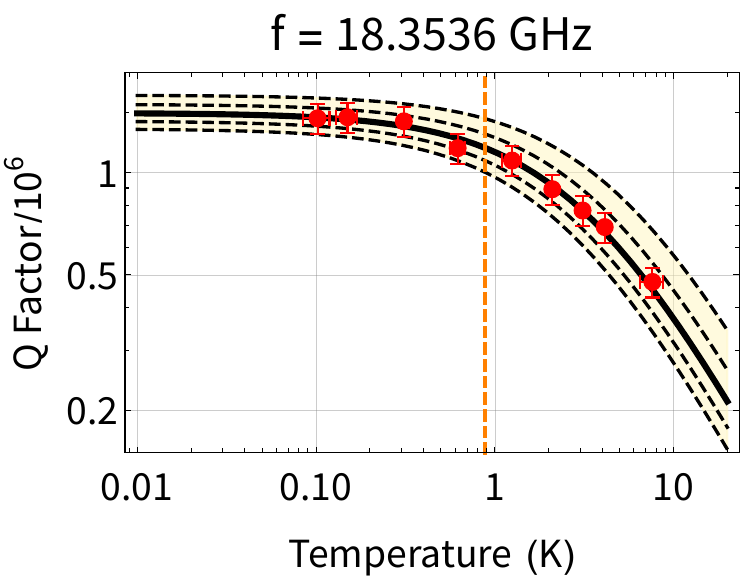}

\centering
\includegraphics[width=0.95\columnwidth]{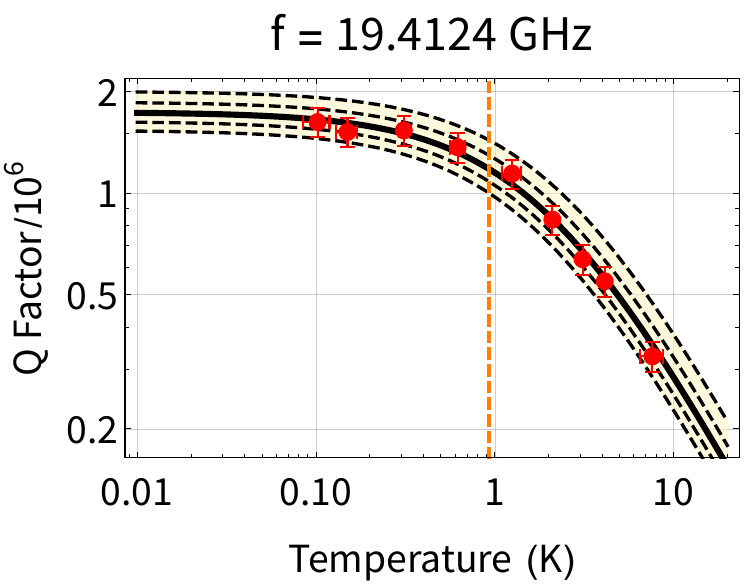}

\centering
\includegraphics[width=0.95\columnwidth]{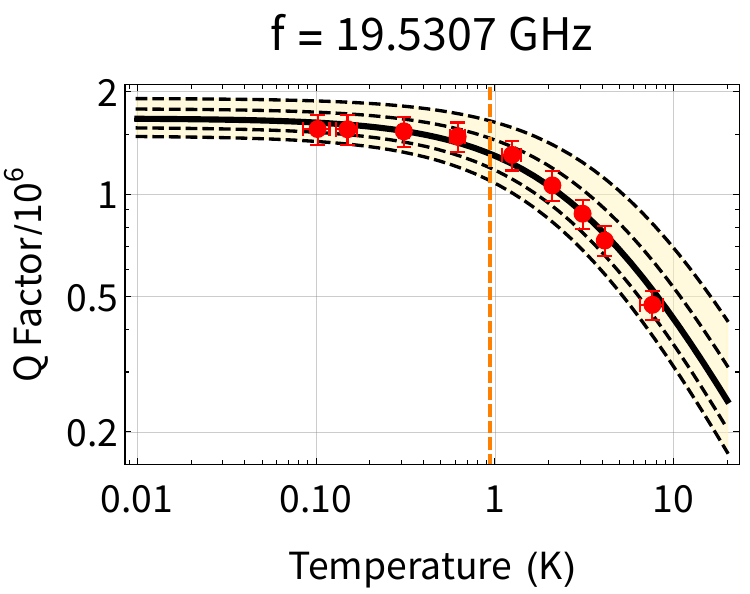}

\centering
\includegraphics[width=0.95\columnwidth]{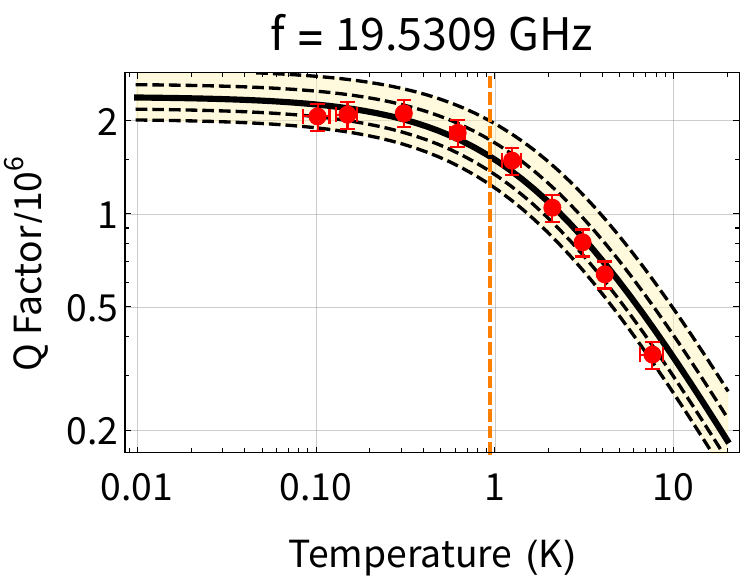}

\centering
\includegraphics[width=0.95\columnwidth]{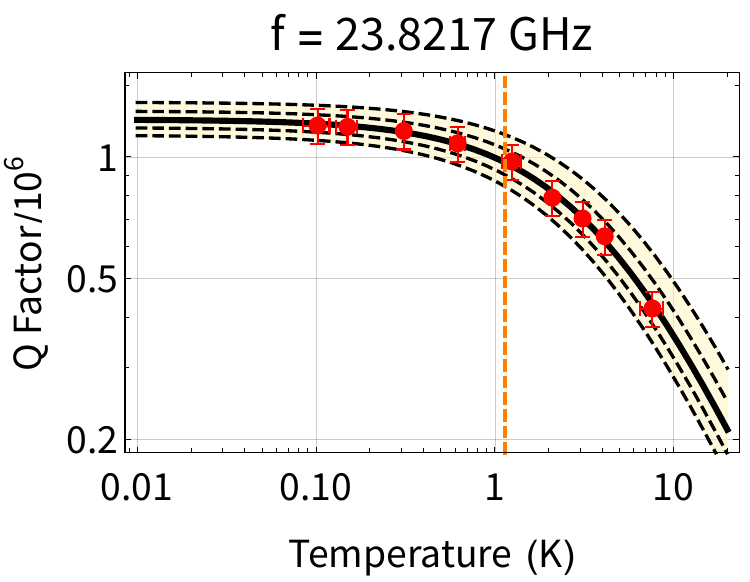}

\centering
\includegraphics[width=0.95\columnwidth]{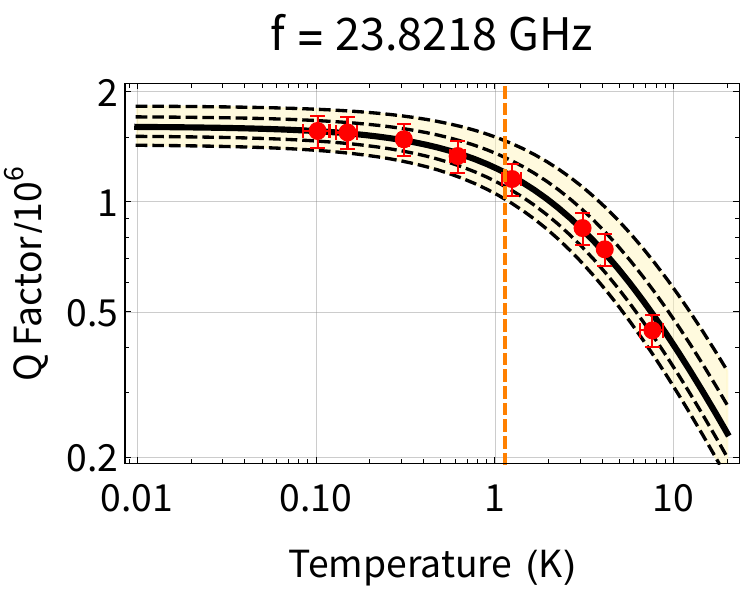}

\end{document}